\documentclass[
reprint,
amsmath,amssymb,
aps,
prl,
]{revtex4-2}

\usepackage{float}
\usepackage{mathtools}
\usepackage{graphicx}
\usepackage{dcolumn}
\usepackage{bm}
\usepackage{upgreek}
\usepackage{lipsum, babel, xcolor}
\usepackage{amsmath, amssymb}
\usepackage{comment}
\usepackage{natbib}
\usepackage{physics}
\usepackage{hyperref}
\hypersetup{colorlinks=true,citecolor=blue}
\usepackage{amsmath}
\begin{document}


\title{Performance of Capacitively Loaded RF Electrodes on SiO$_2$/Si Substrates for High-Speed Electro-Optic Modulators}

\author{Ayed Al Sayem, Ting-Chen Hu, Mark Cappuzzo, Alaric Tate, Rose Kopf, Mark Earnshaw}

\affiliation{%
  Nokia Bell Labs, NJ, USA 
}%
\date{\today}

\begin{abstract}
We experimentally investigate conventional and capacitively loaded traveling-wave RF electrodes fabricated on thermally grown SiO$_2$ on high-resistivity silicon substrates for high-speed electro-optic modulators. The capacitively loaded electrodes exhibit RF attenuation coefficients as low as approximately $0.23\,\mathrm{dB/cm/\sqrt{GHz}}$, comparable to values reported for electrodes fabricated on insulating quartz substrates, while retaining the silicon handle and avoiding substrate removal or undercut processing. By varying the SiO$_2$ buried-oxide (BOX) thickness and the capacitive-loading geometry, the microwave effective index can be engineered over a broad range while maintaining low RF propagation loss. This enables microwave--optical velocity matching over a wide range of optical wavelengths, from the ultraviolet to the telecom C-band. These results demonstrate that thermally oxidized silicon provides a simple and scalable substrate platform for low-loss traveling-wave electrodes in thin-film lithium niobate and thin-film lithium tantalate electro-optic modulators, as well as for high-speed RF interconnects in integrated transceivers and co-packaged optical systems.
\end{abstract}

\maketitle

\section{Introduction}

High-speed, low-voltage modulators are a primary requirement for modern optical interconnects~\cite{Netherton2024,Witzens2018,Miller2009}. A low drive voltage ($V_{\pi}$) is essential for reducing transmitter power consumption and simplifying the electronic driver circuitry ~\cite{Wang2018}. With lower $V_{\pi}$ it is more practical to operate closer to full modulation depth that improves the optical signal quality and bit-error-rate performance of the communication system. Therefore, electro-optic modulators must simultaneously provide a low drive voltage and a large modulation bandwidth~\cite{Wang2018,He2019,Xu2020}. The two major photonic-integration platforms currently used for optical communications and data-center interconnects are indium phosphide (InP) and silicon photonics~\cite{Smit2019InPIntegration,Shekhar2024SiliconPhotonicsRoadmap}. InP provides high-performance active components, including efficient lasers, optical amplifiers, photodetectors, and electro-optic modulators \cite{Smit2019InPIntegration}. However, InP fabrication generally relies on smaller wafers and specialized foundries, making high-volume manufacturing more difficult and expensive than CMOS-compatible silicon-photonics manufacturing~\cite{Smit2019InPIntegration,Rahim2021SiliconModulators}. Moreover, photonic packaging for InP remains costly and has limited manufacturing throughput~\cite{Ranno2022PhotonicPackaging}. Silicon photonics, in contrast, benefits from fabrication on $200$-$300\,\mathrm{mm}$ wafers using mature CMOS manufacturing infrastructure, allowing dense integration and potentially lower costs at high production volumes~\cite{Rahim2021SiliconModulators}. Nevertheless, conventional silicon modulators based on the plasma-dispersion effect are constrained by fundamental trade-offs among modulation efficiency, optical loss, bandwidth, drive voltage, and device footprint~\cite{Rahim2021SiliconModulators}. Consequently, the development of silicon-photonic transceivers is increasingly focused on dense wavelength-division multiplexing, large-scale photonic integration, co-packaged optics using microring modulators, although such modulators require intensive thermal management,
tight fabrication tolerances, and are subject to bandwidth limitations ~\cite{Shekhar2024SiliconPhotonicsRoadmap,Rahim2021SiliconModulators}. Thin-film lithium niobate and thin-film lithium tantalate are promising electro-optic material platforms that provide an attractive combination of low optical loss, strong Pockels nonlinearity, low drive voltage, high linearity, and large modulation bandwidth, with performance competitive with or exceeding that of established silicon- and InP-based modulators in several key metrics~\cite{Zhang2021IntegratedLN,Wang2024TFLT}. 
\begin{figure*}[ht]
    \centering
    \includegraphics[width = 0.85\textwidth]{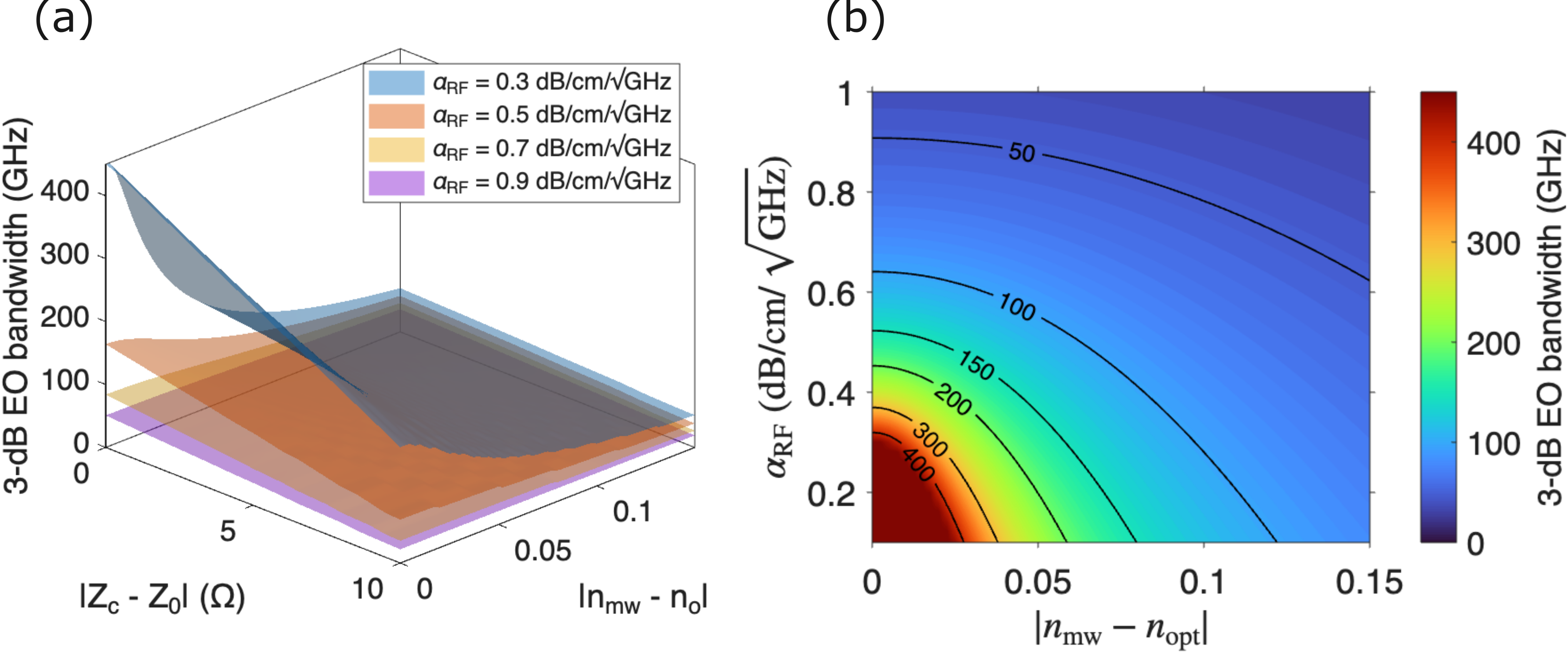}
  \caption{Calculated 3-dB electro-optic (EO) bandwidth of a traveling-wave modulator as a function of RF impedance mismatch, microwave--optical velocity mismatch, and RF propagation loss for a length of 1\,cm. (a) Three-dimensional EO-bandwidth surfaces as functions of the absolute impedance mismatch, $\lvert Z_{\mathrm{c}}-Z_{0}\rvert$, and the absolute index mismatch, $\lvert n_{\mathrm{mw}}-n_{\mathrm{opt}}\rvert$, for $\alpha_{\mathrm{RF}}=0.3$, 0.5, 0.7, and $\mathrm{0.9\,dB/cm/\sqrt{GHz}}$. (b) Contour map of the EO bandwidth as a function of $\lvert n_{\mathrm{mw}}-n_{\mathrm{opt}}\rvert$ and $\alpha_{\mathrm{RF}}$ under impedance-matched conditions. The black contour lines indicate constant 3-dB EO bandwidth in GHz. 
  }  \label{Fig1}
\end{figure*}
For traveling-wave electro-optic modulators based on thin-film lithium niobate (TFLN) or thin-film lithium tantalate (TFLT), the electro-optic bandwidth is determined by impedance matching, microwave-optical velocity matching, and the RF propagation loss of the electrodes~\cite{Zhu2021IntegratedLN,Kharel2021VoltageBandwidth}. Once the impedance and velocity mismatches are properly managed, electrode-induced RF loss often becomes the dominant limitation~\cite{Kharel2021VoltageBandwidth,Chen2022CLTWE}. Increasing the electrode phase shifter length reduces the half-wave voltage, but it also increases the accumulated RF attenuation and can therefore reduce the electro-optic bandwidth~\cite{Zhang2021IntegratedLN,Kharel2021VoltageBandwidth}. Two important approaches for mitigating this limitation are the use of capacitively loaded traveling-wave electrodes on low-permittivity substrates such as quartz~\cite{Tang2025QuartzCLTWE} and the removal or undercutting of the silicon handle beneath oxide-on-silicon devices~\cite{Xu2022SubstrateRemoved,Chen2022CLTWE}. Both approaches, however, have practical limitations. Quartz substrates are not ideal for thermal management, which is a major requirement for both standalone transceivers and co-packaged optics, where heat dissipation is already a serious bottleneck. Quartz also requires significant changes in processing conditions compared to more standard silicon wafers. The silicon-undercut approach adds wafer and chip handling complexity as the undercut regions must remain free of particles, adhesives, and other contaminants to ensure proper device operation. In addition, because LN and LT are strongly piezoelectric materials, microwave modes can readily couple to mechanical modes, introducing either broadband or resonant loss channels and making the devices sensitive to mechanical vibrations. Most microwave-to-optical transducers which operate with the same working principle as electro-optic modulators have avoided this approach \cite{Xu2021BidirectionalTFLN,Holzgrafe2020Cavity,McKenna2020Cryogenic}. Fabricating such structures with high yield and the highest performance can be challenging.

We investigate whether quartz or sapphire substrates, or silicon substrates with an undercut, are necessary to achieve low RF loss. We show that very low-loss RF waveguides can be achieved using high-quality thermal oxide grown on silicon which serves as a BOX layer for both LN and LT. The oxide thickness primarily determines the microwave phase velocity and is therefore important for matching the RF velocity to the optical group velocity. In particular, we demonstrate that low RF loss can be maintained even with a thick BOX layer, while covering a broad range of microwave effective indices required to match the optical group index from the ultraviolet to the C-band. The paper is structured as follows. We first analyze how the electro-optic bandwidth is limited by impedance mismatch, microwave-optical velocity mismatch, and RF propagation loss. We then present detailed RF-loss measurements of planar and capacitively loaded electrodes fabricated on thermally grown oxide layers of different thicknesses on high-resistivity silicon substrates showing the electrode loss reduction achievable with capacitive loading. Next, we examine the corresponding microwave phase indices and demonstrate close to ideal velocity matching with appropriate selection of the BOX thickness. 

\begin{figure*}[ht]
    \centering
    \includegraphics[width = 0.85\textwidth]{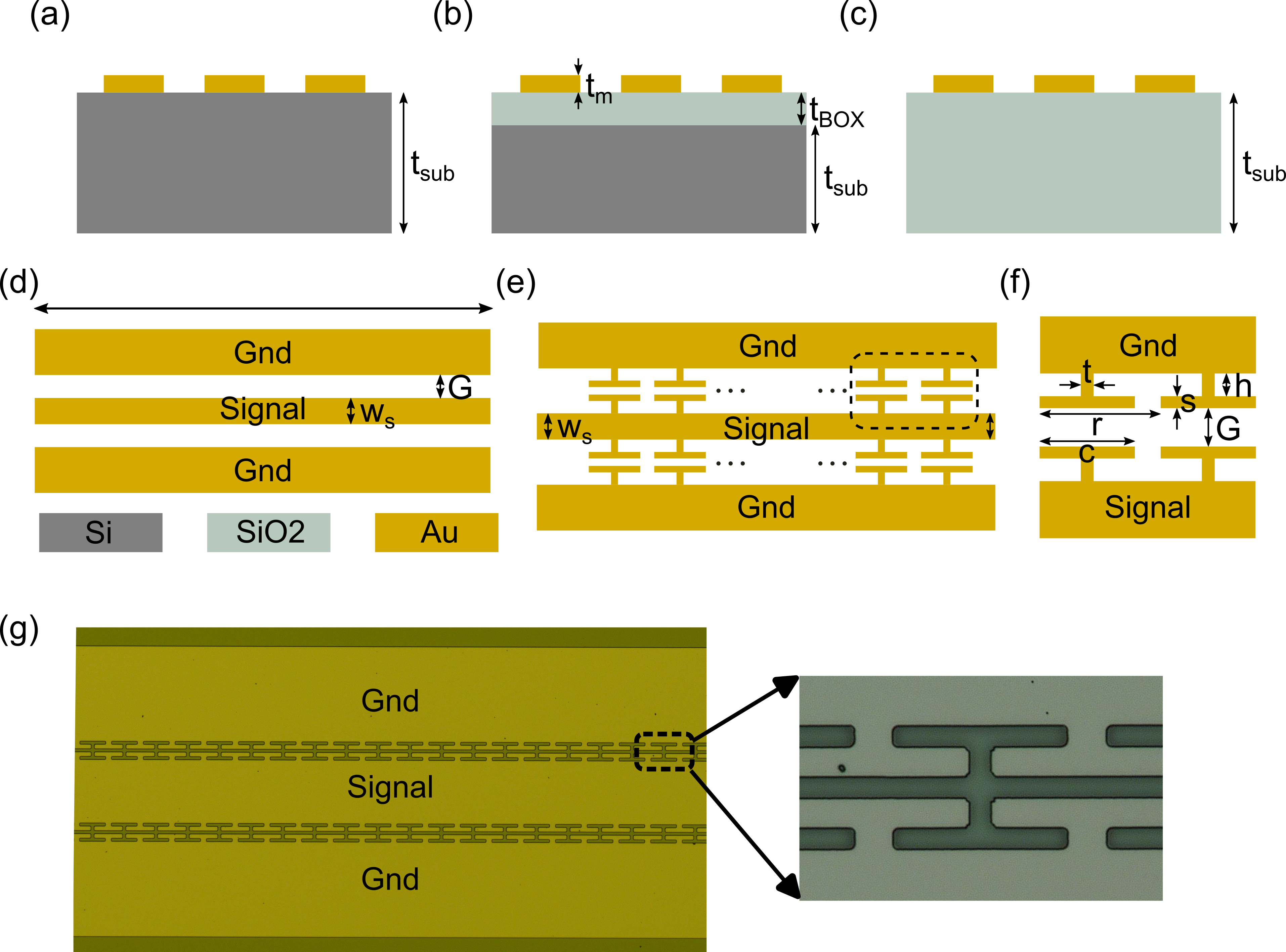}
    \caption{Schematic and optical microscopic image of the substrate configurations and electrode geometries considered in this work. Cross-sectional views of Au electrodes  on (a) Si, (b) a $\mathrm{SiO_2}$-on-Si substrate, and (c) a bulk $\mathrm{SiO_2}$ substrate. The metal and oxide thicknesses are indicated in (b). Top views of (d) a conventional coplanar-waveguide electrode with signal-electrode width $w_s$, (e) a periodically loaded slow-wave electrode, and (f) an enlarged view of the unit cell highlighted by the dashed box in (e). The parameters $t$, $s$, $r$, $h$ and $c$ define the geometrical dimensions of the periodic loading structure. (g) Optical microscope images of the fabricated slow-wave electrode and an enlarged view of its periodic loading structure. The colors corresponding to Si, $\mathrm{SiO_2}$, and Au are indicated in the inset of (d).}
    \label{Fig2}
\end{figure*}

\section{Bandwidth limitation}
We first examine the factors that limit the bandwidth of a traveling-wave modulator on TFLT or TFLN. The EO bandwidth is primarily determined by the impedance mismatch, the phase mismatch between the microwave and optical waves, and the RF propagation loss of the electrodes arising from different loss mechanisms. In Fig.\ref{Fig1}(a), we plot the calculated bandwidth of a TFLT traveling-wave modulator as a function of the absolute impedance mismatch, $\lvert Z_{\mathrm{c}}-Z_{0}\rvert$, and the microwave-optical index mismatch, $\lvert\Delta n\rvert=\lvert n_{\mathrm{mw}}-n_{\mathrm{opt}}\rvert$, for different values of RF loss, $\alpha_{\mathrm{RF}}$. Impedance matching is critical for achieving high bandwidth; however, the impedance of the modulator does not necessarily need to be matched to $50\,\Omega$. Most experimental demonstrations to date have used modulators designed for a characteristic impedance of $50\,\Omega$, primarily because most commercially available characterization equipment operates with a $50\,\Omega$ impedance. In a practical transceiver, however, the modulator impedance must be matched to that of the driver and terminated using a matched on-chip resistance. Therefore, the driver and modulator can be co-designed to achieve optimal performance. 

In this paper, we primarily focus on RF propagation loss and velocity mismatch, as these parameters depend solely on the modulator design. Fig.\ref{Fig1}(b) shows the calculated $3$-dB EO bandwidth as a function of the RF propagation-loss coefficient, $\alpha_{\mathrm{RF}}$, and the absolute velocity mismatch, $\lvert n_{\mathrm{mw}}-n_{\mathrm{opt}}\rvert$, assuming perfect impedance matching. The EO bandwidth decreases as either the RF loss or the velocity mismatch increases. The highest bandwidth is obtained when both parameters are minimized, while even a moderate increase in either parameter significantly reduces the achievable bandwidth. The contour lines indicate constant-bandwidth combinations and illustrate the trade-off between RF loss and velocity mismatch. In particular, a larger velocity mismatch can only be tolerated when the RF propagation loss is sufficiently low, emphasizing that simultaneous reduction of both effects is necessary to achieve ultrahigh-bandwidth operation. RF loss is a major limitation for both high-speed TFLN modulators and microwave--optical transducers~\cite{Kharel2021VoltageBandwidth,Zhang2021IntegratedLN,Holzgrafe2020Cavity,McKenna2020Cryogenic}.  In traveling-wave modulators, increasing the electrode length reduces $V_{\pi}$ but increases RF attenuation, thereby limiting the electro-optic bandwidth~\cite{Kharel2021VoltageBandwidth,Zhang2021IntegratedLN}.  

\begin{figure*}[ht]
    \centering
    \includegraphics[width = 0.85\textwidth]{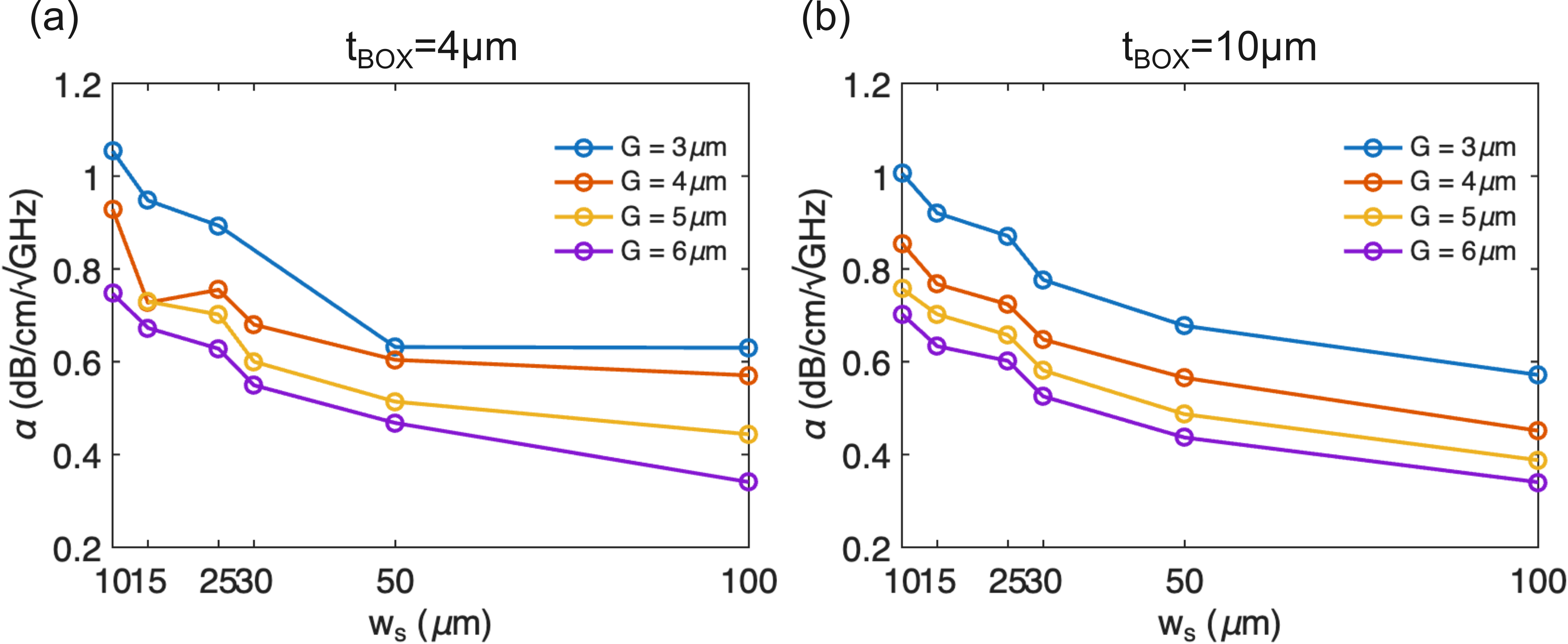}
    \caption{Extracted characteristic RF-loss coefficient, $\alpha$, as a function of the signal-electrode width, $w_s$, for different electrode gaps of $G$ for conventional CPW electrodes as shown in Fig.\ref{Fig2}(b) and Fig.\ref{Fig2}(d). Panels (a) and (b) correspond to BOX thicknesses of $\mathrm{t_{BOX}=4\,\mu m}$ and $\mathrm{t_{BOX}=10\,\mu m}$, respectively.}
    \label{Fig3}
\end{figure*}

\begin{figure*}[ht]
    \centering
    \includegraphics[width = 0.85\textwidth]{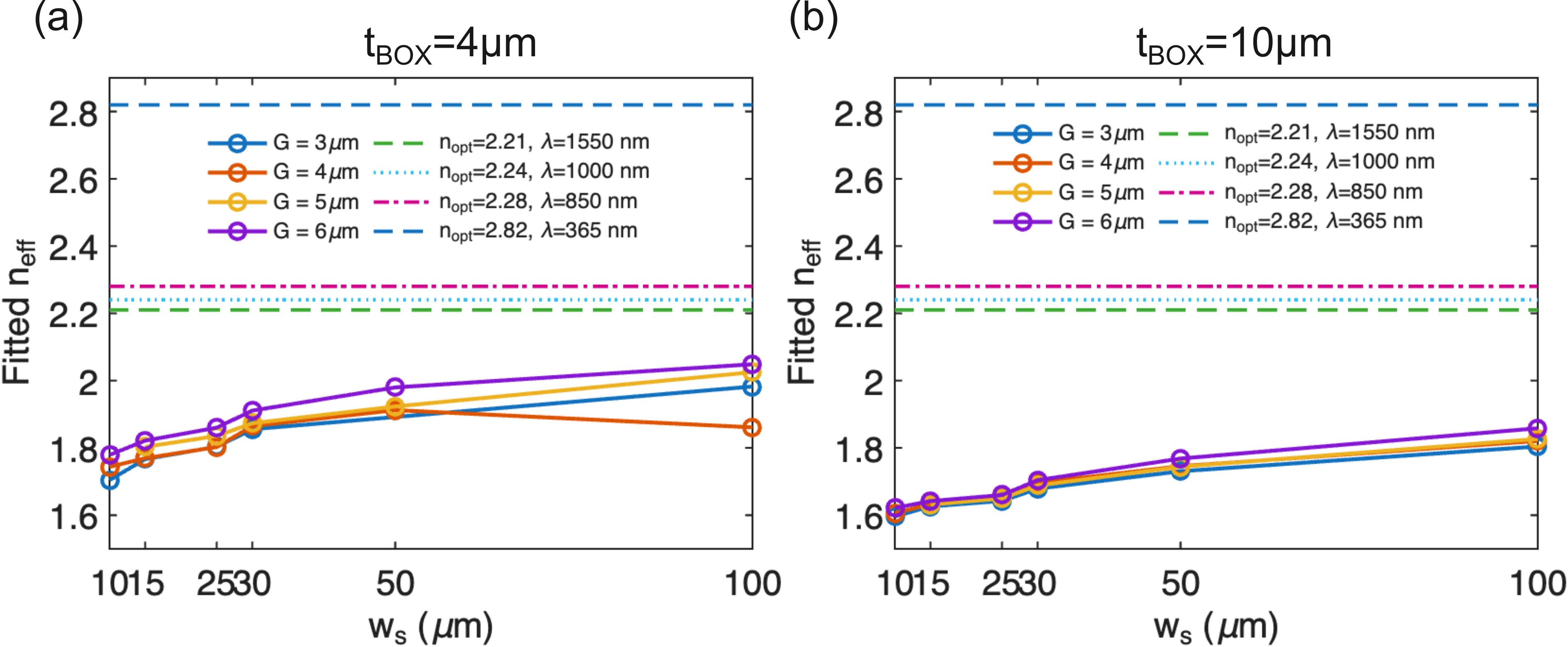}
    \caption{Extracted microwave effective index, $n_{\mathrm{eff}}$, as a function of the signal-electrode width, $w_s$, for electrode gaps ranging from $G=3\,\mu\mathrm{m}$ to $G=6\,\mu\mathrm{m}$. Panels (a) and (b) correspond to BOX thicknesses of $t_{\mathrm{BOX}}=4\,\mu\mathrm{m}$ and $t_{\mathrm{BOX}}=10\,\mu\mathrm{m}$, respectively. The horizontal lines indicate the optical group indices at wavelengths of $1550\,\mathrm{nm}$, $1000\,\mathrm{nm}$, and $850\,\mathrm{nm}$,  $365\,\mathrm{nm}$, providing a comparison of the microwave and optical phase-matching conditions.}
    \label{Fig4}
\end{figure*}

\begin{figure*}[ht]
    \centering
    \includegraphics[width = 0.95\textwidth]{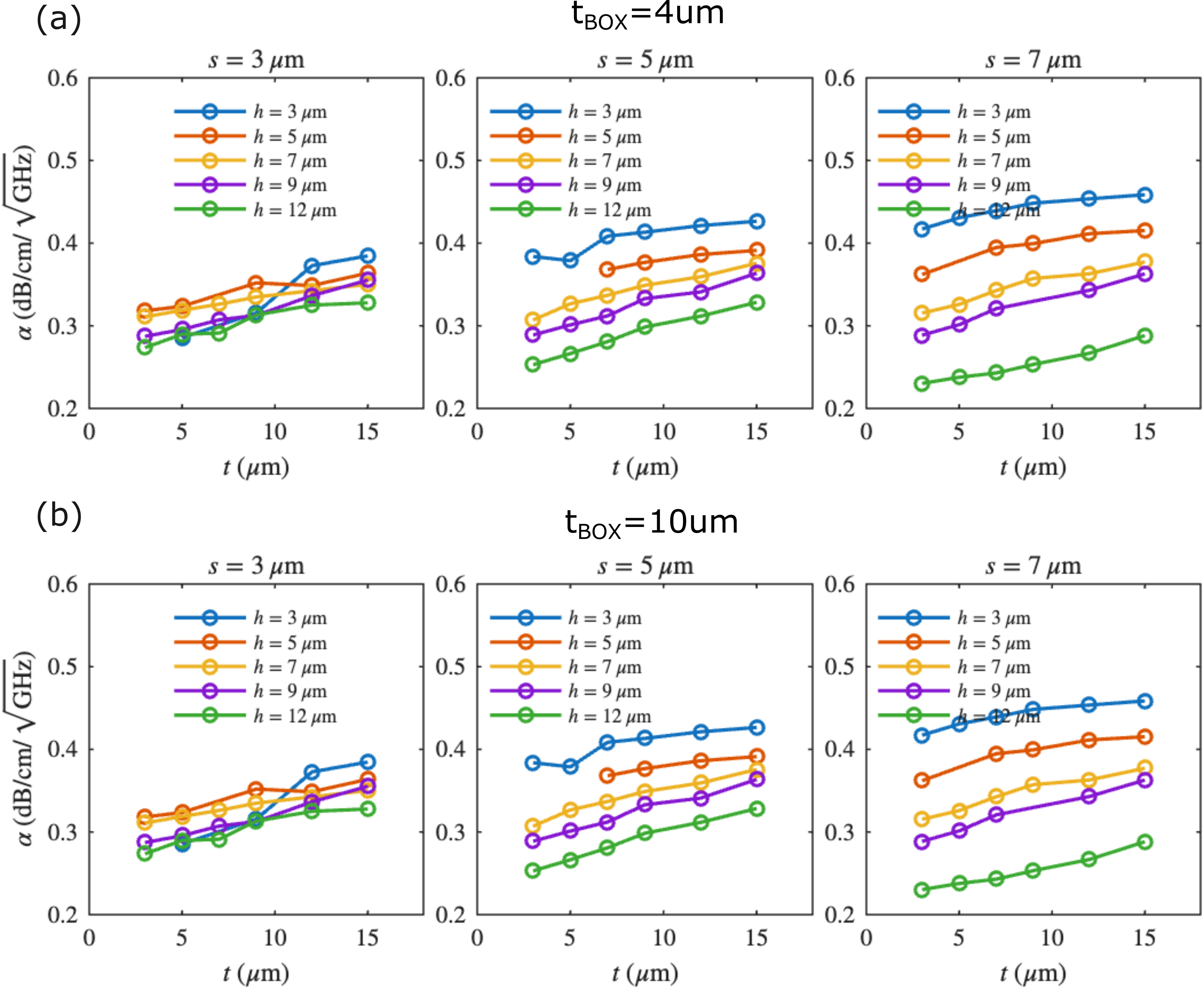}
    \caption{Extracted RF attenuation coefficient, $\alpha$, as a function of the parameter $t$ for different electrode geometries. The three columns correspond to signal-to-ground spacings of $s=3\,\mu\mathrm{m}$, $5\,\mu\mathrm{m}$, and $7\,\mu\mathrm{m}$, while the different curves represent electrode heights of $h=3\,\mu\mathrm{m}$, $5\,\mu\mathrm{m}$, $7\,\mu\mathrm{m}$, $9\,\mu\mathrm{m}$, and $12\,\mu\mathrm{m}$. (a) and (b) correspond to devices with a $4\,\mu\mathrm{m}$ BOX and a length of $10\,\mathrm{mm}$, and a $10\,\mu\mathrm{m}$ BOX and a length of $10\,\mathrm{mm}$, respectively.}
    \label{Fig5}
\end{figure*}

\section{Device design and fabrication}

In this section, we describe in detail the design of the TW-CPW electrodes presented in this paper. Conventional electrodes typically consist of a ground-signal-ground (GSG) geometry. For X-cut TFLN or TFLT modulators, the electrode electric field must be aligned with the Z-axis of the crystal to achieve maximum modulation efficiency. Fig.\ref{Fig2} shows the typical electrode and substrate configurations for a CPW. Figs.\ref{Fig2}(a)-(c) illustrate the cross-sectional geometries of electrodes fabricated directly on a Si substrate, on a SiO$_2$/Si substrate with a variable oxide thickness, and on a fully insulating SiO$_2$ or quartz substrate, respectively. Typical modulators on the TFLN and TFLT platforms use either a Si substrate with a SiO$_2$ BOX layer or a fully insulating SiO$_2$ or quartz substrate. Fig.\ref{Fig2}(d) and Fig.\ref{Fig2}(e) show a conventional coplanar-waveguide electrode and a capacitively loaded slow-wave electrode, respectively. Fig.\ref{Fig2}(f) shows an enlarged schematic of the unit cell, with the geometric parameters $t$, $s$, $r$, $h$, and $c$, which control the capacitive loading and microwave propagation characteristics. Fig.\ref{Fig2}(g) shows an optical microscope image of one of the fabricated devices and a zoomed-in view of the unit cell. The devices are fabricated on thermally grown oxide layers formed on high-resistivity silicon substrates. The electrodes are defined using photolithography and fabricated by electron-beam (E-beam) evaporation followed by a lift-off process. The deposited gold layer has a thickness of $t_{m}=1\,\mu\mathrm{m}$.

\section{Measurement results}
\subsection{RF loss characterization}
In this work, we focus exclusively on experimentally extracted results instead of any simulation. All devices are measured under identical conditions. Symmetric RF probes are used, and the losses from both the probes and cables are removed using standard substrate calibration techniques. We measure the two-port S-parameters, $S_{11}$, $S_{12}$, $S_{21}$, and $S_{22}$. To characterize the RF loss, either $S_{12}$ or $S_{21}$ is fitted as a function of $\sqrt{f}$, from which the RF propagation-loss coefficient, $\alpha$, is extracted. Details of the RF-loss extraction process can be found in the supplementary information. First, we analyze the RF loss and index of a conventional CPW electrode, as shown in Fig.\ref{Fig2}(d). Fig.\ref{Fig3}(a) and Fig.\ref{Fig3}(b) show the fitted RF loss coefficient, $\alpha$, as a function of the signal-electrode width, $w_s$, for different electrode gaps, $G$, and BOX oxide thicknesses of $t_{\mathrm{BOX}}=4\,\mu\mathrm{m}$ and $t_{\mathrm{BOX}}=10\,\mu\mathrm{m}$, respectively. As shown in Fig.\ref{Fig3}, wider electrode gaps and wider signal electrodes reduce the RF loss. However, efficient electro-optic modulation requires a smaller electrode gap to enhance the overlap between the RF electric field and the optical mode, thereby increasing the modulation efficiency. In addition, as described earlier, achieving a high bandwidth requires matching the propagation velocities of the RF and optical modes. In Fig.\ref{Fig4}(a) and \ref{Fig4}(b), we plot the fitted phase index of the RF mode as a function of the signal-electrode width, $w_s$, for different electrode gaps, $G$, and BOX oxide thicknesses of $t_{\mathrm{BOX}}=4\,\mu\mathrm{m}$ and $t_{\mathrm{BOX}}=10\,\mu\mathrm{m}$, respectively. In an actual modulator, the presence of an LN or LT slab and additional oxide cladding can enable velocity matching.

\begin{figure*}[ht]
    \centering
    \includegraphics[width = 0.85\textwidth]{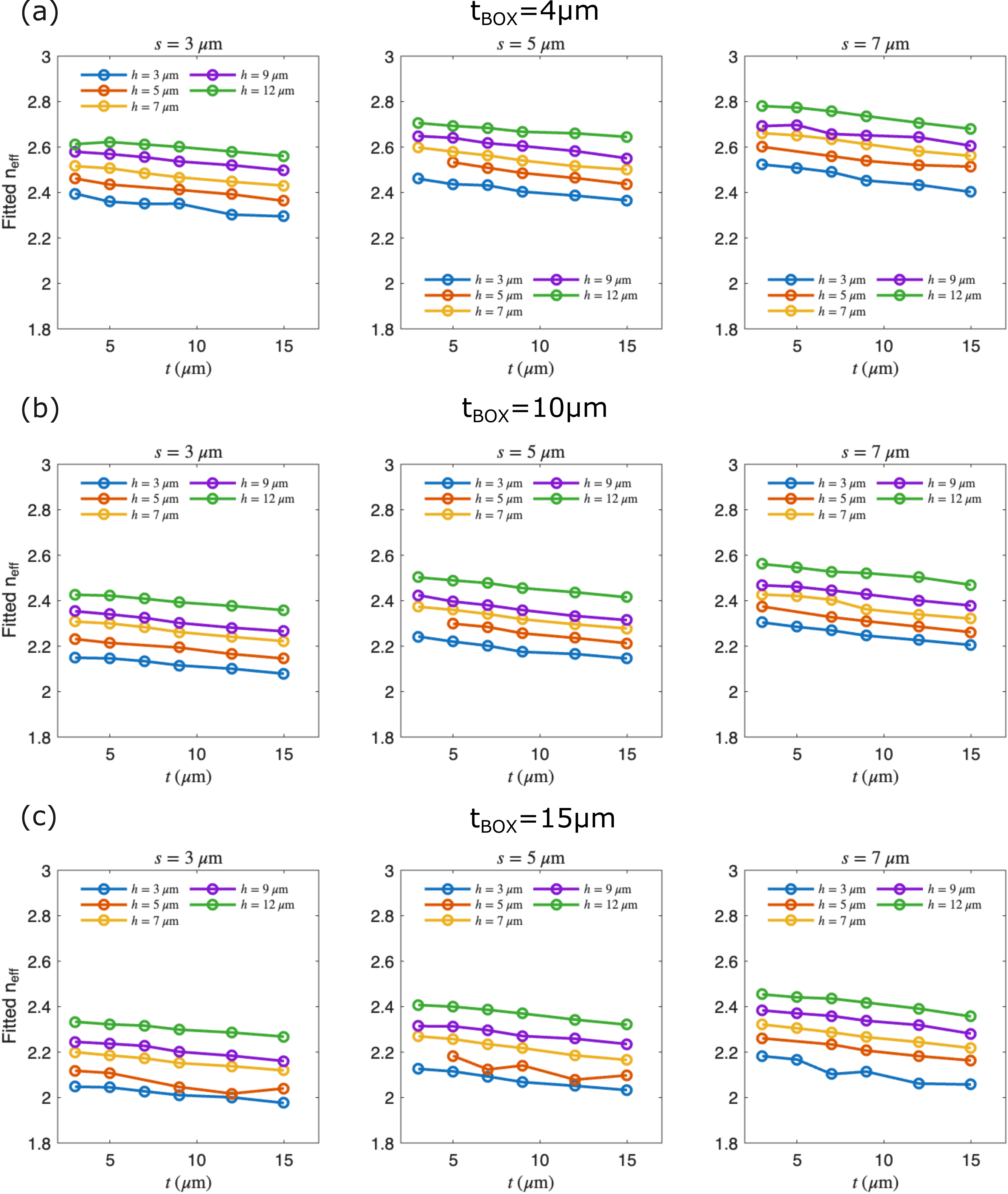}
   \caption{Extracted RF effective index, $n_{\mathrm{eff}}$, as a function of the capacitive-loading parameter $t$ for different values of $s$ and $h$. The three columns correspond to $s=3\,\mu\mathrm{m}$, $s=5\,\mu\mathrm{m}$, and $s=7\,\mu\mathrm{m}$, while the individual curves represent $h=3\,\mu\mathrm{m}$, $5\,\mu\mathrm{m}$, $7\,\mu\mathrm{m}$, $9\,\mu\mathrm{m}$, and $12\,\mu\mathrm{m}$. Panels (a)--(c) correspond to BOX thicknesses of $4\,\mu\mathrm{m}$, $10\,\mu\mathrm{m}$, and $15\,\mu\mathrm{m}$, respectively.}
    \label{Fig6}
\end{figure*}
Next, we focus on the performance of capacitively loaded RF electrodes fabricated on oxide-on-silicon wafers with varying oxide thicknesses. Fig.\ref{Fig5}(a) and Fig.\ref{Fig5}(b) show the extracted RF-loss coefficient, $\alpha$, as a function of the capacitive-loading parameter $t$ for different combinations of $s$ and $h$. In each row, the three columns correspond to $s=3\,\mu\mathrm{m}$, $5\,\mu\mathrm{m}$, and $7\,\mu\mathrm{m}$, while the individual curves represent different values of $h$. In general, $\alpha$ increases with $t$, indicating that stronger capacitive loading results in greater RF propagation loss. The magnitude of the loss also increases with $s$, with the largest values observed for $s=7\,\mu\mathrm{m}$. For fixed values of $s$ and $t$, increasing $h$ generally reduces the RF loss, although minor deviations from this trend are observed for some geometries. The same overall behavior is observed for different BOX thicknesses, demonstrating that the dependence of RF loss on the electrode geometry remains consistent across different thicknesses for the BOX oxides. As shown in Fig.\ref{Fig5}, very low RF loss can be achieved over a broad range of device geometries for different BOX thicknesses. The measured RF loss is comparable to previously reported results for devices fabricated on quartz substrates, which are known to provide some of the lowest RF losses for electro-optic modulators \cite{Kharel2021VoltageBandwidth}.

\subsection{RF phase characterization}

In the earlier section, we showed that similar levels of RF loss can be achieved with different oxide thicknesses over a wide range of parameters. Hence, oxide-on-silicon wafers are suitable for ultra-high-speed modulators. However, achieving high-speed operation requires careful phase matching, especially for long modulators as described in earlier sections. The BOX oxide thickness, $\mathrm{t_{BOX}}$ plays a critical role in matching the microwave phase index to the optical group index. Figure~\ref{Fig6} shows the fitted RF effective index, $n_{\mathrm{eff}}$, as a function of the capacitive-loading parameter $t$ for different combinations of $s$ and $h$. In each row, the three columns correspond to $s=3\,\mu\mathrm{m}$, $5\,\mu\mathrm{m}$, and $7\,\mu\mathrm{m}$, while the individual curves represent different values of $h$. In general, $n_{\mathrm{eff}}$ increases with both $s$ and $h$, indicating that stronger capacitive loading produces a larger RF effective index and, therefore, a lower RF phase velocity. For a fixed combination of $s$ and $h$, the effective index generally decreases as $t$ increases. The same overall dependence is observed for all three BOX thicknesses shown in Figs.\ref{Fig6}(a)--(c), although the absolute value of $n_{\mathrm{eff}}$ varies among the different configurations. Minor nonmonotonic variations observed for some geometries are attributed to experimental fitting uncertainty. In Fig.\ref{Fig4}, we plot the optical group indices of the LT waveguides at the wavelengths of interest. Although low RF loss can be achieved for different oxide thicknesses, a thicker BOX oxide is required to properly velocity-match the microwave and optical waves. The required BOX oxide thickness and the choice of geometric parameters depend on the operating wavelength. However, Fig.\ref{Fig5} and Fig.\ref{Fig6} show that thicker BOX oxides combined with capacitively loaded electrodes are suitable for achieving low RF loss and phase matching over a wide range of optical wavelengths, from the ultraviolet to the infrared, including the C-band.

\section{Conclusion}
In conclusion, we systematically investigate the RF loss of conventional CPW and capacitively loaded electrodes fabricated on SiO$_2$-on-Si substrates and demonstrate that very low RF loss and proper phase-matching conditions over a wide range of optical wavelengths can be achieved using a single electrode layer and a simple fabrication process. Such low-loss electrodes can support not only high-speed EO modulators but also simple, high-speed electrical interconnects between different components, including RF drivers, amplifiers, and modulators, particularly for co-packaged optics.

\section{Author contribution}
A.S. conceived and planned the experiment. A.S. designed the circuits. T.H., M.C., A.T.,  and R.K. fabricated the devices. A.S. performed the measurements and analyzed the data. A.S. wrote the paper with feedback from M.E.

\section{Funding}
Nokia Corporation of America.

\bibliography{Reference}

\end{document}